\documentclass[twocolumn,showpacs,preprintnumbers,pre]{revtex4}
\usepackage{graphicx}
\usepackage{dcolumn}
\usepackage{bm}
\def\b{\begin}
\def\e{\end}
\def\t{thebibliography}
\def\bi{\bibitem}
\def\be{\b{equation}}
\def\ee{\e{equation}}

\def\re{\ref}

\def\r{\rho}

\def\tr{\mbox{Tr}}

\begin{document}

\title{STATISTICAL SEPARABILITY AND THE CONSISTENCY BETWEEN QUANTUM
THEORY, RELATIVITY AND THE CAUSALITY }

\author{Qi-Ren Zhang
\\CCAST(World Lab),P.O.Box 8730,Beijing,100080\\and\\
Department of Technical Physics, Peking University , Beijing,100871,
China\thanks{Mailing address.}}

\vskip0.3cm

\begin{abstract}
We show that the non-locality together with the statistical
character makes the world statistically separable. The super-luminal
signal transmission is impossible. The quantum theory is therefore
consistent with the relativity and the causality.

\end{abstract}

\maketitle

In 1935, Einstein $et\ al$\cite{e} assumed that the world is
separable to prove  the incompleteness of quantum theory. The
separability here means one can always separate two systems by a
space, so that a disturbance act on one of them cannot influence
another system immediately. This separability seemed  to be
self-evident, since otherwise the prompt signal transmission would
be possible. About thirty years later, Bell\cite{b} changed the
thought experiment used in \cite{e} to be a kind of experiment which
can be really done. After that, many experimental results on the
problem of separability have been reported. They show that
Einstein's  assumption of separability is false. One may see this
conclusion in examples \cite{l,t}. The world is proven to be
non-separable in Einstein's sense. This result raises the question
on the consistency between the quantum theory, relativity, and the
causality, and on the possibility of the super-luminal transmission
of signals.

The quantum state of a system is usually defined at a given time but
in the whole space, therefore is non-local and non-separable, in
contrast to the classical state.  Two parts of a system even-though
separated by empty space may be correlated. Their states may be
entangled if they had contacted and therefore interacted each other
previously. People tried to use this entanglement for signal
transmission, and have got experimental success\cite{bo}. It is
called quantum communication. However the super-luminal signal
transmission is still not realized, and it has been proven to be
impossible\cite{g}. It makes people feel that the quantum theory may
be consistent with relativity and causality. In the following we
show that this feeling is reliable.

To make the theory be Lorentz invariant, the quantum state is
defined on a space-like super-surface of space-time in general. It
is non-local. However, the dynamics of the state evolution along a
time-like direction is governed by differential equation. It is
therefore local. According to the multi-time
formulation\cite{d}-\cite{j} of the quantum field theory, the state
$|\sigma\rangle$ on a space-like super-surface $\sigma$ and the
state $|\sigma^\prime\rangle$ on a nearby space-like super-surface
$\sigma^\prime$ are related by the Schr\"odinger equation \be {\rm
i}\hbar\frac{\partial|\sigma\rangle}{\partial\sigma}=-T_{\mu\nu}n_\mu
n_\nu|\sigma\rangle\; .\label{1}\ee Here, the differential ${\rm
d}\sigma$ denotes the infinitesimal 4-dimensional space-time volume
between super-surfaces $\sigma^\prime$ and $\sigma$ around a
space-time point. $(T_{\mu\nu})$ is the energy-momentum tensor
operator of the system and $(n_\mu )$ is the time-like unit normal
4-vector of the super-surface $\sigma$, both at this point. The
solution of (\ref{1}) may be written in the form \be |\sigma\rangle
=U(\sigma ,\sigma_0 )|\sigma_0\rangle\; ,\label{2}\ee with \be {\rm
i}\hbar\frac{\partial
U(\sigma,\sigma_0)}{\partial\sigma}=-T_{\mu\nu}n_\mu n_\nu
U(\sigma,\sigma_0)\; ,\label{3}\ee $\sigma_0$ is an arbitrarily
fixed space-like super-surface. The differential equation shows the
evolution of the state from point to point. The relativity and
causality require that a disturbance of the state $|\sigma_0\rangle$
at the point $({\textit{\textbf{r}}}_0,t_0)$ on it may influence the
properties of state $|\sigma\rangle$ at those points on it only,
which are inside the light cone with its vertex at
$({\textit{\textbf{r}}}_0,t_0)$. Properties of the state
$|\sigma\rangle$ at the point $({\textit{\textbf{r}}},t)$ on
$\sigma$ outside this light cone do not respond to this disturbance.
$U(\sigma ,\sigma_0 )$ is therefore a matrix (or an integral
operator), only those rows and columns have non-zero non-diagonal
elements, which are related to the points inside this light cone.
The remaining part of the matrix (or the integral operator) is unit.

A state of the system is determined by a complete measurement. For a
macroscopic system, the measurement is usually incomplete. In this
case, a density operator, instead of a state, is determined. The
density operator is also defined on a space-like super-surface. Its
time evolution is governed by the von Neumann equation \be
\frac{{\rm d}\r}{{\rm d}\sigma}=-c\left[T_{\mu \nu},\r\right]n_\mu
n_\nu\;\;\; ,\label{4} \ee in which $[A,B]\equiv (AB-BA)/{\rm
i}\hbar$ is the quantum Poisson bracket of $A$ and $B$. The solution
of this equation is \be \r (\sigma )=U(\sigma,\sigma_0)\r
(\sigma_0)U(\sigma_0,\sigma)\;\;\; .\label{5}\ee

Since the super-surface $\sigma$ is space-like, dynamical variables
on different points of it commute with each other. They belong to
different degrees of freedom. Taking the direct products of the
eigenstates of commuting dynamical variables at different points on
$\sigma$ as bases, one may expand the state of the system. Although
the bases in the state are entangled in general, one can still
measure the local property of the system. The local property of the
system at the point $\textit{\textbf{r}}$ on $\sigma$ is described
by the reduced density operator \be \r
(\textit{\textbf{r}})=\tr_{\bar{\textit{\textbf{r}}}}\r(\sigma)\; .
\label{6}\ee The subscript $\bar{\textit{\textbf{r}}}$ in (\ref{6})
denotes that the trace is a sum of matrix elements which are
diagonal respect to the degrees of freedom other than those at the
point $\textit{\textbf{r}}$ only. We show in the following that if
$({\textit{\textbf{r}}},t)$ is outside the light cone of the point
$({\textit{\textbf{r}}}_0,t_0)$, the disturbance at
$({\textit{\textbf{r}}}_0,t_0)$ cannot influence the the reduced
density matrix $\r ({\textit{\textbf{r}}},t)$, therefore cannot
influence the result of the local measurement at point
$({\textit{\textbf{r}}},t)$. We call this property the statistical
separability.

Denote the degrees of freedom in a macroscopically infinitesimal
neighborhood of the point $\textit{\textbf{r}}$ by $a$, and all
other degrees of freedom by $b$. The bases to be used in expanding
the state of the system are
$[|n_a,n_b\rangle\equiv|n_a\rangle|n_b\rangle]$. $n_a$ is a complete
set of quantum numbers for degrees $a$ of freedom, while $n_b$ is
that for degrees $b$ of freedom.  In this representation, we may
write \be \r (\sigma_0)\equiv \sum_{n_a,n_b,n_a^\prime ,n_b^\prime
}|n_a^\prime\rangle n_b^\prime\rangle\r_{n_a^\prime n_b^\prime ;\,
n_an_b}\langle n_a| \langle n_b|\; ,\label{7}\ee $[\r_{n_a^\prime
n_b^\prime ;\, n_an_b}]$ are the matrix elements of $\r (\sigma_0)$.
According to the argument after (\re{3}), if
$({\textit{\textbf{r}}},t)$ is outside the light cone of the point
$({\textit{\textbf{r}}}_0,t_0)$, we may write \be
U(\sigma,\sigma_0)\equiv \sum_{n_a,n_b,n_b^\prime }|n_a\rangle
|n_b^\prime\rangle U_{n_b^\prime ,\, n_b}\langle n_a| \langle n_b|\;
,\label{8}\ee and \be U(\sigma_0,\sigma)\equiv
\sum_{n_a,n_b,n_b^\prime }|n_a\rangle |n_b^\prime\rangle U_{n_b
,\,n_b^\prime }^*\langle n_a| \langle n_b|\; .\label{9}\ee The
relation $U(\sigma_0,\sigma)U(\sigma,\sigma_0)=1$ requires \be
\sum_{n_b^\prime}U_{n_b^\prime n_b^{\prime\prime}}U_{n_b^\prime
n_b}=\delta_{n_b^{\prime\prime},n_b}\; .\label{a}\ee Substituting
(\ref{7}-\ref{9}) into (\ref{5}), then substituting the result into
(\ref{6}), by use of (\ref{a}) and the ortho-normality of base
states $[|n_a\rangle|n_b\rangle]$ we obtain \b{eqnarray} \r
(\textit{\textbf{r}})&=& \sum_{n_an_a^\prime n_bn_b^\prime
n_b^{\prime\prime}}|n_a^\prime\rangle
U_{n_b,n_b^{\prime\prime}}\r_{n_a^\prime
n_b^{\prime\prime};\,n_an_b^\prime}U^*_{n_b,n_b^\prime}\langle
n_a|\nonumber\\&=&\sum_{n_an_a^\prime
n_b^\prime}|n_a^\prime\rangle\r_{n_a^\prime
n_b^\prime;\,n_an_b^\prime} \langle n_a|=\r_0
(\textit{\textbf{r}})\; ,\label{10}\e{eqnarray} in which \be \r_0
(\textit{\textbf{r}})\equiv\sum_{n_an_a^\prime
n_b}|n_a^\prime\rangle\r_{n_a^\prime n_b;\,n_an_b} \langle
n_a|\equiv \tr_{\bar{\textit{\textbf{r}}}}\r(\sigma_0)\ee is the
reduced density operator at point $\textit{\textbf{r}}$ on the
super-surface $\sigma_0$. Since the local reduced density operator
is the complete description of the local measurement in quantum
theory, (\ref{10}) shows that the result of local measurement at a
point does not change until the point goes into the light cone of
the disturbance. This is the statistical separability of the space.
One may immediately conclude that although the quantum states are
non-local and local states at two points separated by space-like
distance may be entangled (non-separable), the communication between
two space-likely  separated points (super-luminal communication) is
still impossible by disturbance and measurement on a quantum system.
The causality in relativity is therefore  ensured by the present
quantum theory.

The argument for the possibility of super-luminal communication by
use of the non-locality of the quantum state is based on an
assumption that the relation between the disturbance and the
measurement result is deterministic. This is not true in quantum
theory. According to quantum theory, disturbance changes the quantum
state, and therefore changes the statistical distribution of the
measurement results. The deterministic theory with non-local state
contradicts the relativity and the causality. Fortunately, in
quantum theory, one works  not only with non-local state but also
with original statistics. This two wings make the quantum theory be
consistent with relativity and causality.

This work was supported by the National Nature Science Foundation of
China (Grant No.10305001).

\b{\t}{99} \bi{e}A.Einstein, B.Podolsky, and N.Rosen, Phys.Rev. {\bf
47}(1935) 777.
\bi{b}J.S.Bell Physics {\bf 1} (1964) 195.
\bi{l}M.Lamehi-Rachti and W.Mittig, Phys. Rev. D{\bf 14}(1976) 2543.
\bi{t}W.Tittel {\it et. al.} Phys. Rev. Lett. {\bf 81}(1998) 3563.
\bi{bo}D Bouwmeester {\it et. al.} Nature {\bf 390} (1997) 575.
\bi{g}G.C.Ghirardi {\it et. al.} Lett. Nuovo Cimento {\bf 27}(1980)
293.
\bi{d}P.A.M.Dirac, Proc. Roy. Soc. A{\bf 136} (1932) 453.
\bi{s}S.Tomonaga, Progr. Theor. Phys. {\bf 1}(1946) 27.
\bi{j}J.Schwinger, Phys. Rev. {\bf 74} (1948) 1439. \e{\t}

\end{document}